\documentclass[letterpaper,twocolumn,10pt]{article}
\usepackage{usenix-2020-09}
\usepackage{nopageno} 
\usepackage{amsmath,amssymb,amsfonts}
\usepackage{textcomp}
\usepackage{xcolor}
\usepackage{url}
\usepackage{booktabs}
\usepackage{multirow}
\usepackage{comment}
\usepackage{tikz}
\usepackage{arydshln}

\newcommand{\Simley}[1]{%
\begin{tikzpicture}[scale=0.11]
    \newcommand*{\SmileyRadius}{1.0}%
    \draw [fill=brown!10] (0,0) circle (\SmileyRadius)
        ;  

    \pgfmathsetmacro{\eyeX}{0.5*\SmileyRadius*cos(30)}
    \pgfmathsetmacro{\eyeY}{0.5*\SmileyRadius*sin(30)}
    \draw [fill=black,draw=none] (\eyeX,\eyeY) circle (0.15cm);
    \draw [fill=black,draw=none] (-\eyeX,\eyeY) circle (0.15cm);

    \pgfmathsetmacro{\xScale}{2*\eyeX/180}
    \pgfmathsetmacro{\yScale}{1.0*\eyeY}
    \draw[color=black, domain=-\eyeX:\eyeX]   
        plot ({\x},{
            -0.1+#1*0.15 
            -#1*1.75*\yScale*(sin((\x+\eyeX)/\xScale))-\eyeY});
\end{tikzpicture}%
}%

\newcommand{\amii} {\textcolor{black}{\textit{am I infected?}}}

\newcommand{\update}[1] {#1} 

\newcommand{\updatefigC}[0] {}

\newcommand{\updatenumC}[2] {#2} 
\newcommand{\updatenumY}[2] {#2} 

\newcommand{\todo}[1] {#1} 

\begin{document}


\title{Am I Infected? Lessons from Operating a Large-Scale IoT Security Diagnostic Service}

\author{
{\rm Takayuki Sasaki$^1$\textsuperscript{\textsection}, Tomoya Inazawa$^1$\textsuperscript{\textsection}, Youhei Yamaguchi$^1$, Simon Parkin$^{2,1}$,}\\ {\rm Michel van Eeten$^{2,1}$, Katsunari Yoshioka$^1$, and Tsutomu Matsumoto$^1$}\\
\\
{\rm $^1$Yokohama National University, $^2$Delft University of Technology}
}

\maketitle

\begingroup\renewcommand\thefootnote{\textsection}
\footnotetext{Two authors contributed equally to this work.}
\endgroup

\begin{abstract}
There is an expectation that users of home IoT devices will be able to secure those devices, but they may lack information about what they need to do. In February 2022, we launched a web service that scans users' IoT devices to determine how secure they are.
The service aims to diagnose and remediate vulnerabilities and malware infections of IoT devices of Japanese users.
This paper reports on findings from operating this service drawn from three studies: (1) the engagement of \updatenumC{98,722}{114,747} users between February, 2022---\updatenumC{July, 2023}{May, 2024}; (2) a large-scale evaluation survey among service users ($n=4,103$), and; (3) an investigation and targeted survey  ($n=90$) around the remediation actions of users of non-secure devices. 
During the operation, we notified \updatenumC{389~(0.39\%)}{417~(0.36\%)} users that one or more of their devices were detected as vulnerable, and \updatenumC{138~(0.14\%)}{171~(0.15\%)} users that one of their devices was infected with malware. 
The service found no issues for 99\% of users. Still, 96\% of all users evaluated the service positively, most often for it providing reassurance, being free of charge, and short diagnosis time. 
Of the \updatenumC{138}{171} users with malware infections, \updatenumC{49}{\todo{67}} returned to the service later for a new check, with \updatenumC{44}{\todo{59}} showing improvement. Of the \updatenumC{389}{417} users with vulnerable devices, \updatenumC{140}{\todo{151}} users revisited and re-diagnosed, where \updatenumC{69}{\todo{75}} showed improvement. We report on lessons learned, including a consideration of the capabilities that non-expert users will assume of a security scan.

\end{abstract}

\section{Introduction}
Users are exposed to a lot of generic security advice. It is hard to prioritize which of these actions they need to focus on -- leading some experts to observe a ``crisis of advice prioritization'' \cite{redmiles2020comprehensive}. One mechanism to help this prioritization is to tailor the advice to evidence of the actual risks that the specific user faces. Web services have emerged to serve that purpose. Think of web services which allow users to check whether their personal credentials have been breached, such as \href{https://haveibeenpwned.com/}{\textit{haveibeenpwned.com}}~\cite{have-i-been-pwned}; 
IP reputation services that inform users whether their IP addresses are involved in malicious activities such as scanning or spam~\cite{IP-Reputation-Check, IP-Reputation-Check-CISCO};
and services that check exposed TCP services on routers or hosts~\cite{tcp-port-scan, tcp-port-scan2, tcp-port-scan3}.
An alternative approach is to scan via browser extensions (e.g., \cite{Thomas2019ProtectingAccounts}), but they require users to find and install software on their machine, so it is less accessible and requires higher levels of trust than a web service. 

In February 2022, we launched \amii, a web service supporting Japanese users to diagnose and remediate vulnerabilities and malware infections for their IoT devices. Many IoT devices are exposed to the Internet~\cite{shodan, censys, insecurity-of-embedded-devices} and regularly get infected by malware such as Mirai~\cite{understanding-mirai-botnet, circle-of-life}. To mitigate these issues, notification campaigns have been launched to inform owners of vulnerable or compromised devices, either by ISPs~\cite{ccetin2019cleaning} or by the government~\cite{notice}. While these campaigns were effective in encouraging remediation, they rely on entities contacting users. Users cannot seek out this information themselves. Hence, our motivation to launch \amii. It aims to provide feedback on various IoT security issues for the specific user, rather than the single-issue notification campaigns that existed so far.   

To the best of our knowledge, there is little academic research on user engagement and experience with web services to diagnose and remediate user security issues, let alone for IoT devices specifically. How do the users evaluate their experience of using the service? Does it incentivize and enable them to remediate the issues? We present data from \updatenumC{16}{27} months of operation, serving \updatenumC{98,722}{114,747} users; a large-scale user survey ($n = 4,103$) on their evaluation of the service; and an investigation and small survey  ($n = 90$) around the remediation actions of users.
\textcolor{black}{Of course, this population of users is self-selected, namely the users who were willing to visit the service. That said, many security solutions are voluntary, e.g., anti-virus, browser extensions, VPNs, so research on those solutions in real-world settings also faces self-selection issues. We would argue that it is important to evaluate these solutions in order to improve the support for users, even if that group does not include everyone.}
We aim to contribute to remediating IoT security issues and to evaluating web services for tailored security advice as a complementary alternative to notifications via browser extensions, service provider emails or unsolicited emails from third parties~\cite{rodriguez2022difficult, maass2021effective, Comparing-Large-Scale-Notifications, Best-Practices-for-Notification}.

We explore the main lessons around three questions. \textit{RQ1: Is a diagnostic service for IoT devices helpful for general consumers?} It is not clear what users expect from such a service. We conducted a survey to determine whether users felt the service was useful and why -- or why not. The results showed that 96\% of users gave a positive evaluation. In addition, 96\% of users answered that they would like to continue using the service. 
We also asked users about the good and bad points of the service. 
The most frequent response about good points is providing reassurance~(18\% users). The users also mentioned easy diagnosis, a short diagnosis time, and being free of charge.
Regarding the bad points, 60\% of users mentioned that there are no specific issues. 11\% of users highlighted the lack of diagnosis details as bad points. In addition, 8\% of users pointed out the untrustworthiness of the emails, for example, our reminder emails seemed like phishing emails.

\textit{RQ2: How many security issues with IoT devices were detected? And how many of these issues could be mitigated?}
During \updatenumC{16}{27} months of operation, the service conducted \updatenumC{157,086}{195,598} diagnoses initiated by \updatenumC{98,722}{114,747} users. We identified vulnerabilities for \updatenumC{389}{417} users and malware infections for \updatenumC{138}{171} users. We confirmed that \updatenumY{49\%}{\todo{50\%}} of vulnerabilities and \updatenumY{90\%}{\todo{88\%}} of malware infections were remediated.

\textit{RQ3: What are obstacles to remediation for users?}
To clarify the reasons for non-remediation, 
we surveyed service users with security risks by questionnaires. Among users with remediation not confirmed,
we identified that 56\% of them attempted to take measures, but of those users, 17\% did not complete them, mostly due to difficulty in identifying or operating devices with issues. Of the users who did not attempt to take measures, 30\% of users answered that they did not know how to take measures.

In summary, our contributions are:
\begin{itemize}
    \item  We deployed and operated a Web-based security diagnosis service that allows users to diagnose the risks of IoT devices when the users want. During \updatenumC{16}{27} months of operation, we diagnosed \updatenumC{98,722}{114,747} users.  
    \item 
    We find that web services can support remediation, with over half of the users being able to remediate the issue. This means web services can provide an alternative to more general security advice sources~\cite{redmiles2020comprehensive} or notification mechanisms where users are reliant on third parties, such as ISPs, to contact and inform them~\cite{ccetin2019cleaning}.
    \item 
    A follow-up survey of the users with security issues, users found that the obstacle was that users did not know how to act on the recommendations. The users also mentioned that financial cost was an issue when the advice was to replace the IoT device.
\end{itemize}

\section{Web-based diagnosis service}
\label{sec:method}
\label{subsec:method-amii}

Here we detail the goals of the service, its scope, and how it operates for users.

\subsection{Design goals}
\label{sec:design-goals}

\amii~has the following features compared to the existing notification methods via email by ISPs or third parties~\cite{rodriguez2022difficult, maass2021effective, Comparing-Large-Scale-Notifications, Best-Practices-for-Notification}, open port scan service~~\cite{tcp-port-scan, tcp-port-scan2, tcp-port-scan3} dedicated to port scanning for experts, and browser-extension/android-app-based notification~\cite{warpdrive} requiring software installation. 
\update{Specifically, an internal agent would indeed allow more efficient tracking, but would raise adoption barriers.
We chose a web-based service because we do not necessarily need to track users in order to advise them. 
} 

(1) Services available at any time and to anyone: Our service is implemented as a web service accessible through a web browser. Users can have their network scanned, and diagnose risks at any time.

(2) Approaches for non-expert users: To make the service approachable for non-experts, the system provides content regarding the risks and the service.
Specifically, we set up two types of content: content on the top page of the service to recommend diagnosis and content to tell the identified risks and measures.
As for the content on the top page, we present what happens when users leave malware infection and vulnerabilities, for example, infected devices can potentially attack others. In addition, in the FAQ, we provide basic knowledge of IoT security, such as what a `vulnerability' is and what `malware' is.
We created an FAQ, referring to our knowledge, a FAQ page of existing notification project~\cite{notice-faq}, and the government's website that provides security information~\cite{mic-security-web-site}.

As for the content to convey identified risks and measures, we present actionable measures, as shown in Appendix~\ref{sec:notification-messages}. Behaviours were identified which are in line with existing broad guidance, e.g., change default passwords, install updates, check a manufacturer website for instructions. 

(3) Wide range of diagnosis: Our service diagnoses a comprehensive range of security risks and notifies users about those risks. Specifically, in addition to malware infection, our service also notifies the user of known vulnerabilities, the use of old firmware, known initial passwords, etc. Our service diagnoses the risks shown in Table~\ref{tab:diagnositc_items_and_recommended_actions}. 

(4) Minimal cost of entry:
By providing the system as a Web service, users can check the security situation of their IoT without the installation of an application. In addition, our service does not depend on the type of browser or OS.

\subsection{Scope of the service}
The risks of vulnerable consumer IoT devices are discussed worldwide, for example, UK security laws for smart devices~\cite{uk-security-laws}, the EU Cyber Readiness Act~\cite{eu-cyber-resilience-act}, and cybersecurity labeling for consumers in the US~\cite{us-nist-iot-label}. Therefore, the need to engage users of non-secure IoT devices is a broadly international concern.

\update{
In this paper, devices connected to the Internet are broadly defined as IoT devices. Specifically, diagnostics are conducted on the following types of devices that can be examined by a vulnerability scanner~(See Section~\ref{subsec:system-design}.):
router,
NAS,
web camera,
firewall,
WiFi access point,
video recorder,
industrial devices,
home energy management,
home network management, and
printer.
Although the vulnerability scanner supports a wide variety of devices, since the users of this service are general consumers, the identified security issues were for four types~(router, NAS, web camera, firewall)~(Section~\ref{subsec:identified-risks}).
}

Due to the variety of IoT devices, our service currently supports \update{devices frequently used in Japan.} The architecture of our service does not depend on \update{specific} devices.
However, to support international devices, we have to have signatures to identify the devices and information on vulnerabilities of the devices. 
To extend our service to support more devices, collaborations with foreign security organizations would be required.

\subsection{How to use the service}
\label{subsec:how-to-use-amii}
Users need to proactively visit the service. After implementing measures, users revisit the service to conduct a re-diagnosis and check whether the security risks have been mitigated by the measures taken.

\begin{figure}[t]
  \centering
  \includegraphics[keepaspectratio, width=0.85\linewidth]{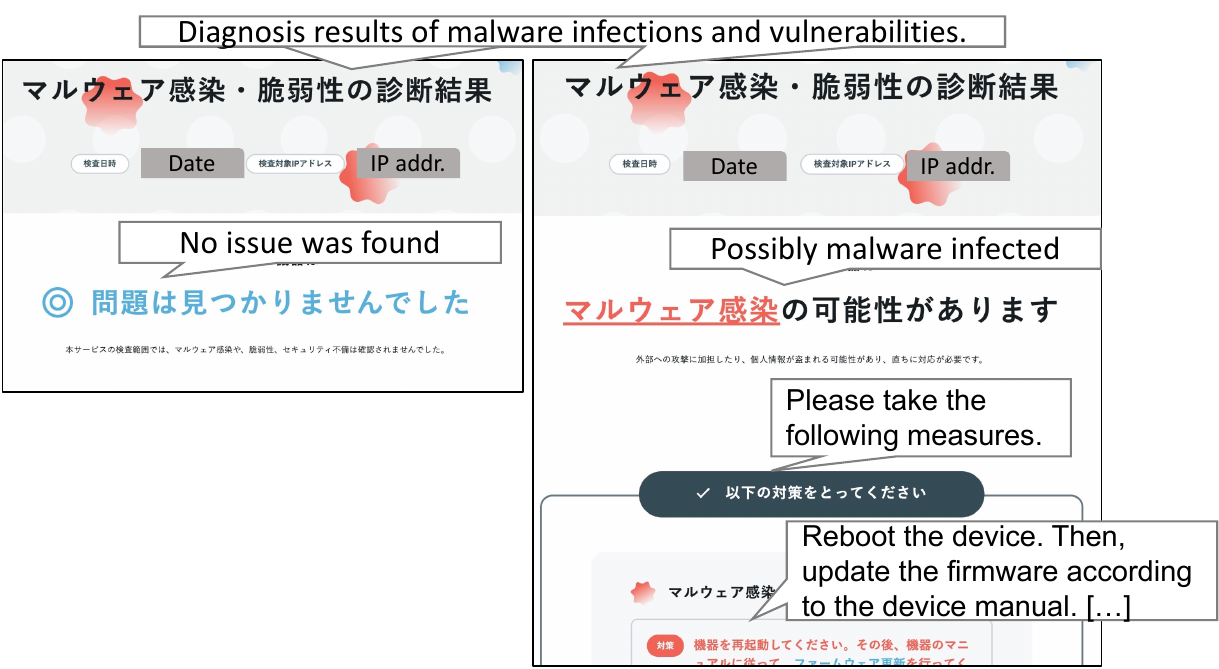}
  \caption{Diagnosis result pages}
  \label{fig:amii-result} 
\end{figure}

\textbf{Step 1: Request diagnosis.}
On the top page, the user enters their e-mail address and indicates their location~(home, workplace, outside, etc.). The user would also log how they heard of the service (Web media article, our university website, lecture at a conference or seminar, Twitter~(X), etc.). The user would then click the ``Start Infection Diagnosis'' button to request a diagnosis.

\textbf{Step 2: Receive diagnosis result.} A few minutes later, the user receives an e-mail notifying them of the completion of the inspection. They can then access a Web page for viewing the inspection results from the URL described in the body of the e-mail. 
Our service detects security risks and shows recommendations as listed in Table~\ref{tab:diagnositc_items_and_recommended_actions}.
Specifically, the notification message on the web page comprises the type of the detected risk, information about the device with the risk~(the manufacturer and model name), and the recommended measure. The left side of Figure~\ref{fig:amii-result} shows an example of the web content when risk is not detected. The right side of Figure~\ref{fig:amii-result} shows an example of the web content when malware infection is detected. Users are recommended to update their firmware. Detailed notification messages translated from Japanese are shown in Appendix~\ref{sec:notification-messages}.

When security issues are found, the service displays a button to request support on the results page. When users click the button, we contact them and provide support via email. This feature was implemented 14 months after the service launched.

\textbf{Step 3: Take measures and re-diagnose.} Users at risk should take countermeasures, and request a re-diagnosis by clicking the re-diagnosis button on the web page of the diagnosis results. Users who have not been re-diagnosed within a certain period\footnote{In the beginning, the period was one week, and after eight months of service, it was reduced to three days. The reason for this was to prompt users while they were still motivated immediately after using the service.} will receive a reminder email to prompt them to take measures and re-diagnose.

\begin{table}[t]
    \centering
    \caption{Scope of diagnosis and recommended measures}
    \label{tab:diagnositc_items_and_recommended_actions}
    \footnotesize	
    \begin{tabular}{l@{\hskip5pt}p{4cm}}
        \\\toprule
        Risk &  Recommendation measure\\ 
        \midrule
        Malware infection & Update firmware and reboot \\
        Risky protocol (Telnet) &Disabling Telnet, Replacement of the IoT device\\
        End of support & Replacement of the IoT device\\
        Admin password not set & Password change\\
        Known vulnerability & Firmware update\\
        Old firmware  & Firmware update\\
        Known ID/credential & ID/password update\\
        Vulnerable default Wi-Fi pass.  & Wi-Fi password change\\
        No authentication  & Replacement of the IoT device if used in critical infrastructure\\
        \bottomrule
    \end{tabular}
\end{table}

\section{Methodology}
\label{section:methodology}

In this section, we describe service design, implementation, deployment, and basic statistics about the users and diagnoses. Then, we explain two user surveys: engagement and obstacles to taking measures.

\subsection{Service implementation and deployment}
\label{subsec:system-design}
Here, we explain the methods to detect the risks of malware infection and vulnerabilities.

\update{
\textbf{Threat model and overview of diagnosis method.}
The threats our service targets are cases where the user is operating a vulnerable device and is at risk of being the target of an attack, such as malware injection, unintended use of the device features, and interception of information. In addition, our service also targets cases where an attack has already taken place and the device is infected with malware. 
Due to the detection methods described below, our service only covers risks that can be scanned from the Internet.
}

\textbf{Detection method of malware infection.}
Our service leverages the honeypots and darknet observation data to determine the presence of malware infection. Specifically, we use a Telnet honeypot using real IoT devices and an HTTP honeypot that can extend imitation of IoT devices by collecting responses of IoT devices from the Internet. Darknet is an observation system using unused IP address ranges.

Our service checks whether the user's IoT device has accessed honeypots and/or the darknet.
\update{Since the detection of malware infection is based on honeypot and darknet observation, we do not identify the device model.}
Specifically, if the user's IP address is present in the observation data from the past 24 hours at the time of inspection, our service determines that the user's IoT device is infected. The reason for restricting the matching to data from the last 24 hours is to prevent false positives caused by dynamic IP address assignments. Because the detection method relies on the latest 24 hours of data, users must wait at least 24 hours from the time of inspection to obtain the correct result for malware infection. The 24-hour duration is a provisional value, and we need to consider the optimal settings based on future surveys of IP churn~\cite{ip-churn1, ip-churn2}. 
It is probably worth noting that malware not continuously propagating or less active may be missed and cause false negatives.

\textbf{Detection method of vulnerability.}
Our service utilizes the IoT scan engine Karma~\cite{karma}, developed by 00One, Inc. 
\update{The Karma's scan engine was deployed on our university's network.}
\update{Note that the vulnerability scanner itself is not part of our contributions.}

\update{Karma uses non-harmful methods, such as collecting banner information and HTML, and does not perform active penetration testing to exploit vulnerabilities. Specifically, }
Karma probes the typical TCP ports used for WebUI~(e.g., TCP 80 and 8080) and receives a response from the device. Karma has device signatures to identify the models of IoT devices based on the responses from the IoT devices~\cite{karma-patent}.

Additionally, Karma has a database that includes vulnerability data, each associated with a specific device model. To identify risks, Karma retrieves responses from a user's IP address and determines the device model by matching the obtained response and device signatures. Once Karma identifies the device \update{model}, it also identifies vulnerabilities of that device model. According to the Karma developer page, it can identify more than 5,000 device models. The detection of the Telnet service can be performed regardless of the identification of the device model, because detection is based on the open/close status of the Telnet port~(port 23/TCP).

Our service notifies the detected vulnerabilities and the corresponding countermeasures based on the results of the on-demand inspection of the user's IP address using Karma. If Karma successfully identifies the relevant device, the vendor, series, and model of the device, our service also provides the information to users.

\update{
\textbf{False positives and false negatives.}
False positives of malware infection can be caused by multiple users sharing the same IP address, such as an ISP line shared by multiple tenants. In addition, detection of malware infection might produce false positives when tested less than 24 hours after measures have been taken. False positives of vulnerabilities can be caused by device fingerprinting errors, such as a device being fingerprinted incorrectly as a previous model or previous~(firmware) version with a vulnerability.
}

\update{
False negatives in malware detection result from malware activities~(scanning and infection spreading) not reaching our honeypot and darknet, even though Mirai variants do randomly traverse the whole public IPv4 address space.  False negatives for vulnerabilities occur when the models of the devices cannot be identified, because vulnerabilities are determined by identifying the model and whether that model has vulnerabilities. 
}

\textbf{Security of the service.}
To prevent a large number of diagnosis requests from a bot, we deployed Google reCAPTCHA~\cite{recapcha} for sending diagnosis requests. We also deployed an access control mechanism for the diagnosis pages using Cookie so that each result page can be accessed only by the user who requested the diagnosis.

\textbf{Service deployment.}
\label{sec:service-deployment}
We deployed and operated our service according to the timeline in Table~\ref{tab:timeline}.
From the service launch on February 24, 2022, to \updatenumC{July 1, 2023}{May 31, 2024}, our service got \updatenumC{98,722}{114,747} users.

\begin{table}[t]
    \centering
    \caption{Service timeline}
    \label{tab:timeline}
    \footnotesize
    \begin{tabular}{@{\hskip2pt}p{2.1cm}@{\hskip5pt}p{6.0cm}@{\hskip2pt}}
    \toprule
    Date & Event\\
    \midrule
    Feb. 24th, 2022& Service launched.\\
    Feb. 26th &  One of us appeared on a TV program of a national TV station, explained the risks of IoT devices, and introduced our service as a free, easy-to-use service that diagnoses vulnerabilities and malware infections. \\
    Mar. 7th & Introduced by a national TV station in the context of how to prepare for cyber attacks.\\
    Mar. 9th - 15th & Under maintenance.\\
    Jun. 15th&  For a survey of obstacles to remediation~(Sec.~\ref{sec:user-survey-barriers-to-take-measures}), questionnaires were sent to users with risks identified between the service launch and June 12th, 2022.\\
    Jun. 22nd & Reminders of the questionnaires were sent to the users who had not yet answered the questionnaires sent on 15th June.  The last response was received on July 19th.\\
    Nov. 24th & We created a Twitter account and posted the user count and users with security issues every week. \\
    May 1st, 2023&  We implemented the questionnaire function on the diagnosis result page for a survey on engagement~(Sec.~\ref{sec:user-survey-procedure}).\\
    May 1st& A button for inquiry support was implemented on the diagnosis result page. \\
    May 10th - 19th & A follow-up survey campaign to recommend re-diagnosis was conducted. Emails were sent to users who had used the service between its launch and May 5th, 2023, to ask them to run the service again.\\
    \bottomrule
    \end{tabular}

\end{table}

\subsection{Participants}
\label{subsec:usage_trends}
We assumed participants learned about the service through Japanese TV programs and newspapers and provided the service in Japanese. However, we did not implement any usage restrictions, such as restricting access from foreign countries.
\update{According to IP address geolocation data~\cite{maxmind-geoip}, 99\% users were from Japan.}
\textcolor{black}{According to a questionnaire on the service's top page from May 1, 2023, to June 1, 42\% of users learned about our service through various media (e.g., TV, newspapers, and online media), 24\% through web searches, and 7\% through social media, with the remaining 18\% through other sources.}

\begin{figure}[t]
    \centering
    \includegraphics[width=\linewidth, trim={4mm 5mm 6mm 0},clip]{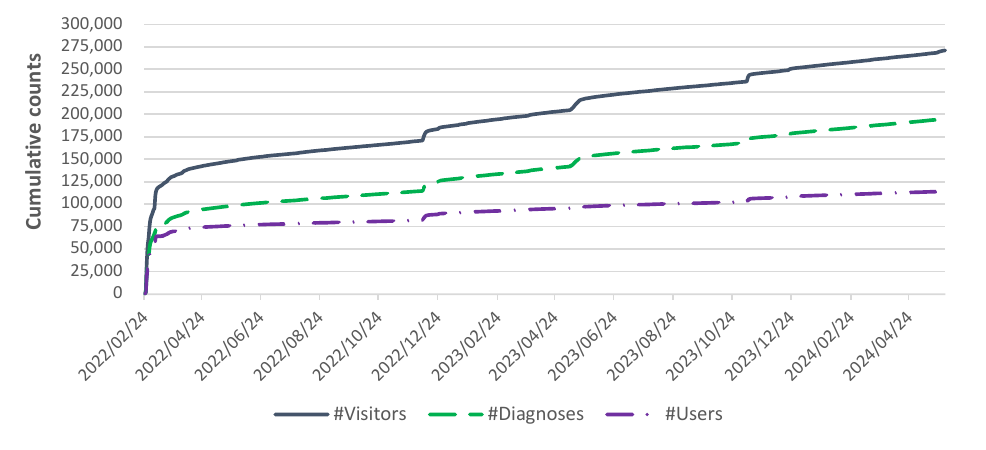}
    \caption{Cumulative number of users and diagnoses.\updatefigC} 
    \label{fig:transition_of_number_of_diagnosis_and_users}
\end{figure}

Figure~\ref{fig:transition_of_number_of_diagnosis_and_users} shows the number of users and diagnoses.
`Visitors' shows unique visitors to our service top page obtained using Google Analytics. `Users' means the number of users who requested diagnoses.
Immediately after the service was launched, the number of users increased rapidly due to the effect of media coverage. Subsequently, the number has been gradually increasing since the maintenance period (March 9 to March 14) due to the service version upgrade. 
In December 2022, there was a second steep increase in users caused by an introduction by a newspaper.
In May 2023, there was a third steep increase in diagnoses caused by a follow-up campaign.
Investigating locations of the \updatenumC{153,028}{191,540} diagnosis requests\footnote{Until May 1, 2023, requesting a re-diagnosis from the results page does not require an answer to the current location, and thus location information is not available for some re-diagnoses.}, \updatenumC{136,283~(89\%)}{169,205~(88\%)} of the requests were for ``home,'' indicating that the service was used as intended. The number of requests for ``workplace'' was \updatenumC{10,832~(7\%)}{14,114~(7\%)}, ``outside the office/home'' was \updatenumC{2,058~(1\%)}{2,961~(2\%)}, and ``other'' was \updatenumC{3,855~(3\%)}{5,260~(3\%)}. 
A user may diagnose at several places such as home and workplace. Specifically, \updatenumC{3\%}{4\%} of the entire service users took diagnoses in multiple places \update{according to questionnaire responses}, and also we confirmed \updatenumC{6\%}{7\%} of entire service users took diagnoses in multiple network environments \update{by analyzing system logs}.
There were \updatenumC{43}{55} users who have taken diagnoses in several environments and risky devices have been detected at least once.
Among those \updatenumC{43}{55} users, \updatenumC{3}{6} users have been detected with risks in 2 different environments.

The majority of users, \updatenumC{70,980~(72\%)}{80,998~(71\%)}, had been diagnosed only once. 
The percentage of users who had been diagnosed more than twice was only \updatenumC{28\%}{29\%}, and users did not use the service continuously. Some users continued to use the service, for example, \updatenumC{537}{880} users \update{with and without the risk} used the service more than 10 times.
With regards to the number of usages by users without the risk, it turned out that \updatenumC{27,466 of 98,198~(28\%)}{33,434 of 114,162~(29\%)} users used more than once. 
With regards to those whose device \update{has the risk}, it was found that \updatenumC{276 of 524~(53\%)}{315 of 585~(54\%)} users used more than once.
To keep IoT devices secure, periodical diagnoses are desirable. However, most users did not revisit our service.

During the operation of the service, we received \updatenumC{94}{104} emails asking questions or requesting support. 
We also received \updatenumC{two}{four} requests via the button to ask for support on the results page.
We discuss the details in Appendix~\ref{sec:individual-support}.

\subsection{Survey on user engagement}
\label{sec:user-survey-procedure}

We implemented a questionnaire form on the diagnosis result page. Users do not have to complete the survey to see the diagnosis results, and the survey is voluntary.

\textbf{Questionnaire items.}
We set up three sets of questionnaires: for users without security issues and for users with the issues before and after taking measures.
Specifically, Table~\ref{tab:questionnaire-items-2nd-survey} shows the questions to ask about the helpfulness of the service and the strength/weakness of the service. For all users, we asked their willingness to continue using the service~(Likert scale from 1 to 5) and the helpfulness of the service~(where this is a 3-item scale with emojis, an approach used elsewhere to capture sentiment in security studies, e.g., \cite{you-still-use-the-password}).
Additionally, we asked for good and bad points about the service; the questions were broad, to minimise the amount of time visitors to the service would be expected to commit to providing responses.
A screenshot of the questionnaire form is in Figure~~\ref{fig:questionnaire-about-engagement} in Appendix~\ref{sec:questionnaire-about-engagement}.
For users with security issues, we inquired whether they were willing to take action and whether they could do so themselves. Once we confirmed that the issue was mitigated, we asked about the measures taken, whether the measures were completed or not, and any challenges encountered in taking those measures.

\begin{table}[t]
    \centering
    \caption{Questionnaire items (Translated from Japanese)
    }
    \label{tab:questionnaire-items-2nd-survey}
    \footnotesize
    \begin{tabular}{@{\hskip2pt}p{1.55cm}p{6.3cm}@{\hskip2pt}}
    \\\toprule
All users & $\bullet$ Would you plan to continue to use this service? (It’s desirable to diagnose your devices regularly as new vulnerabilities are constantly reported and attack trends can change) (Likert scale:1 to 5)*\\
         & $\bullet$ Is this service helpful? (three items with emojis: helpful \Simley{1}, neutral \Simley{0.3}, and not helpful \Simley{-1})\\
         & $\bullet$ Please tell us the good points. (Open-ended) \\
         & $\bullet$ Please tell us the bad points. (Open-ended)\\
         \midrule
\multirow{2}{*}{\shortstack[l]{Users  with\\ issues (before \\taking measures) }}& $\bullet$ Do you wish to take the measures? (Likert scale:1 to 5) \\
      & $\bullet$ Do you think you can take the recommended measures by yourself? (Likert scale:1 to 5) \\
      \midrule
\multirow{2}{*}{\shortstack[l]{Users \\ with issues\\(after taking \\measures)}} & $\bullet$ Measures we proposed: [Recommended measure that the service proposed]. What measures did you consider to take? (Open-ended) \\
    & $\bullet$ Did you complete the measures? (Yes/No) \\
    & $\bullet$ Did you have any challenges in taking measures? (Open-ended) \\
    \bottomrule
    \end{tabular}
    \scriptsize
    \flushleft{* For users with security issues, we asked this question after the issues were remediated.}
\end{table}

\textbf{Coding of questionnaire responses.}
After we received the questionnaire responses, we coded the open-ended responses. We followed the principles of thematic analysis \cite{braun2006using} \cite{braun2021one}, with codes being generated inductively by two coders and through regular codebook meetings.
Specifically, two researchers started by reading through the first 200 responses and generated an initial codebook. We generated codes nested with different levels of abstraction. For example, if a user responded ``It's trustworthy,'' a code good-trustworthiness was generated, and if a user responded with a reason, such as ``A national university operates it, so it's trustworthy'' a code, good-trustworthiness-of-site-operator, was generated. Then, the coding was conducted by the two researchers in two steps.

\textit{Code assignment:} Each researcher assigns one or more codes to the answers. Multiple codes were assigned if a single response contained more than one comment on the good or bad points of the service.

\textit{Consistency check and codebook update: }If there is an inconsistency in code assignment between the two researchers, we discuss and resolve the inconsistency by consolidating, adding, or removing codes. Discussions were held on 23\% of the code assignment results for the first 500 responses, and as the codebook got fixed, the rate decreased to about 14\% for the last approximately 400 responses.

For responses from 4,103 individuals, we conducted the above steps for every 500 responses, and codes were discussed among the author group. 
Since our dataset is large, we employed simple automation for responses regarding bad points. Specifically, most users provided identical responses, such as ``nothing,'' ``nothing in particular,'' or ``no bad points,'' so we assigned the code using exact match rules. For good points, there was a variety of responses, such that automated code assignment was not viable.
As for good points, 77\% of the answers had one code, 19\% had two codes, and the remaining had 3-5 codes. 
As for bad points, 96\% of the answers had one code, and 4\% had two codes. The number of responses with 3 and 4 codes was only 17.

\subsection{User survey: obstacles to taking measures}
\label{sec:user-survey-barriers-to-take-measures}

In order to clarify the actual status of users who were notified of the risks, we conducted a follow-up survey of 367 users whose risks were detected. 

E-mails requesting the questionnaire survey were sent to 367 users with identified risks, and 90 users~(25\%) completed the questionnaire. 
We divided the users into two groups: (A) a user group whose remediation was confirmed by re-diagnosis and (B) a user group whose remediation was not confirmed~(Users who still have the risks or who did not re-diagnose). 

For both groups, we defined questions about whether or not the users took the measures. 
For users whose remediation was not confirmed, there were two possibilities: they gave up in the middle of the remediation steps, or they took measures but did not re-diagnose.
For users who give up on remediation, according to a prior notification study~\cite{user_compliance_and_remediation} as a reference, we assumed four steps: (1) understanding the necessity of remediation and understanding how to take measures, (2) identification of the device with risks, (3) taking the measure to the device, and (4) completing the measures. According to the steps, we designed the questionnaire to clarify where the users stumbled between steps (1) and (4). Detailed questionnaire items are shown in Figure~\ref{fig:questionnaire_items_for_A} and \ref{fig:questionnaire_items_for_B} in Appendix~\ref{sec:questionnaire-items}.

Once the questions were designed, we used a Google Form to create a multiple-choice questionnaire. We created four types of questionnaires in order to adjust the wording of the question items and the answer choices depending on whether the detected risk was a malware infection or a vulnerability. Specifically, we sent a questionnaire based on whether the identified risk was a vulnerability or a malware infection and whether the device was remediated or not.

Finally, we created e-mails including the date and time of diagnosis, the diagnosis result, and a link to the questionnaire. Then, we distributed the e-mails to the email address that users entered when the users requested diagnoses. The e-mail was sent on June 15th, 2022, and was resent on June 22nd to users who had not responded. 
When we sent questionnaires, 14 non-remediated users were misgrouped as users with remediation confirmed due to multiple diagnoses at several environments\footnote{A user performed diagnosis in an environment with a risk, then the user performed diagnosis at another environment without the risk. We were not aware of that problem at the time we sent the questionnaires. 
}. We excluded the data of the 14 users from the analysis.

\subsection{Ethical considerations}

This research project was approved by our institution's ethics review board~(IRB). Additionally, the deployment of the service was approved by an official meeting with the Chief Information Security Officer~(CISO) of our institution.

The privacy policy of our service was hosted on the website \textcolor{black}{as a part of the terms of service}, and it stated that the collected information is used for our research. We also specified that anonymized data will be published as an academic paper. 
\textcolor{black}{The diagnosis request button is labeled ``I agree to the terms of use and start diagnosis,'' prompting the user to agree to the terms of use. The words ``terms of use'' are hyperlinked to the text so that users can read it by clicking on them.}
Furthermore, when we received a request to delete information collected by our service, we deleted the information related to the user in our database.
\update{Regarding the questionnaires sent to users via email, we implemented a mechanism to unsubscribe and explained how to unsubscribe in the email body.}


\section{Results: user survey}

\begin{table}[t]
    \centering
    \caption{Statistics of questionnaire responses} 
    \footnotesize
    \label{tab:statistics-of-questionnaire-responses}
    \begin{tabular}{@{\hskip3pt}l@{\hskip8pt}l@{\hskip5pt}p{2.1cm}@{\hskip3pt}p{2.7cm}@{\hskip3pt}}
        \\\toprule
& Clean users & Users with issues & Users who re-diagnosed \\ 
\midrule
\#Targets & 11,431 & 18 & 5 \\
\#Respondents & 4,100 & 3 & 1 \\
\#Responses & 4,371 & 3 & 1 \\
        \bottomrule
    \end{tabular}
\end{table}

Here we answer RQ1, assessing user response to the service. We got responses from 4,100 clean users and 3 users with security issues \update{ between May 1st and July 1st, 2023}~(Table~\ref{tab:statistics-of-questionnaire-responses}).

Of the users without security issues, 96\% of users positively rated the service as helpful~(users selected helpful out of a 3-item scale: helpful, neutral, and not helpful).
96\% answered that they would be willing to continue using the service~(4 or 5 out of 1-5 Likert scale). This indicates that users were very satisfied with the service. 
Of 3 users with security issues, all of them rated this service as helpful.

As for good and bad points, we generated 29 codes for good points and 28 codes for bad points and assigned the codes to responses. 
Usefulness is a theme related to how helpful the diagnostic service was to the user; usability is a theme related to how user-friendly the diagnostic procedure was, and trustworthiness is a theme related to the trustworthiness of the service and results. In the following sections, we conduct a detailed analysis of the responses based on these themes.

\begin{figure*}[t]
    \centering
\includegraphics[width=0.95\linewidth, trim={3mm 2mm 10mm 3mm},clip]{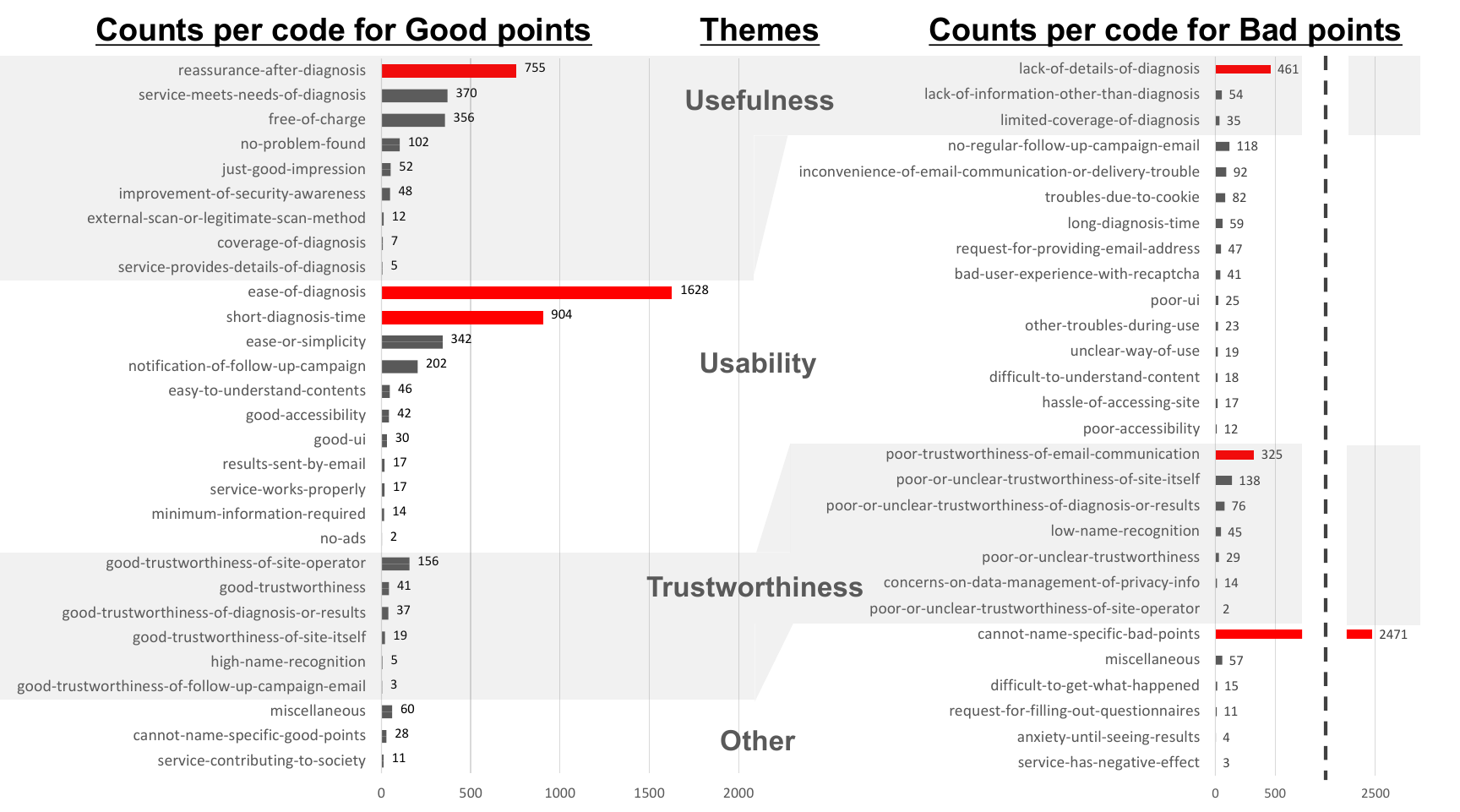}
    \caption{
    Coding results of good and points about the service
    (red bars indicate the top 3 codes for good and bad points)
    }
    \label{fig:good-bad-points}
\end{figure*}

\subsection{Representative responses on good points}
\label{sec:response-good-points}
\textbf{Summary}
The left part of Figure~\ref{fig:good-bad-points} displays the coding results of good points. We identified that users mentioned ease of diagnosis~(n=1628) a short diagnosis time~(n=904), reassurance after diagnosis~(n=755), free of charge~(n=356) as good points. 

\textbf{Reassurance~(Usefulness)}
Most users feel reassurance after use of the service~(code:reassurance-after-diagnosis, n=755), for example, \textit{``Thanks to multiple security checks, I feel more reassurance.''}
This result means that even if security issues are not detected, the service is meaningful to the user for providing reassurance. 

Traits similar to ``fear appeals''~\cite{renaud2019cyber} may have led to the opinion that they were reassured. Our service and its promotion explained the risks of IoT devices, which may have prompted users to act, and within this created a sense of urgency to act (much like fear appeals). 
However, according to the survey results, 37 respondents stated that they had concerns before using the service (pointing to a service like this one addressing an existing concern
among the population of device users). The remaining users did not clearly state when they started feeling anxious.

Similarly, several users pointed out that having no problems found was a good point~(code:no-problem-found, n=102). For example,\textit{ ``It's good to know that the device is secure for now.'' }
We are unsure if users feel reassured by the diagnosis results, but knowing that things are safe is meaningful for them even when their devices are indicated as having no detectable problems.

\textbf{Free of charge~(Usefulness)}
Users noticed that free of charge is a good point~(code:free-of-charge, n=356). For example, \textit{``Since I did not want to install paid software to find out if my device was infected by malware, a free service that I could easily check was ideal for me right now. [..]''}
It is clear that the service being free of charge is one of the factors that make users feel comfortable using the service. 
Some users were grateful for being able to use a service for free that could reasonably be paid for. For example, \textit{``[..] It is also good that it is free for now. However, I am willing to pay a fee for this service.''}
In comparison, other work has found that points of contact in retail feel it necessary to ensure that security protections are available to match user budgets, including free options for those who have not budgeted for security \cite{parkin2019security}.

\textbf{Matching user needs~(Usefulness)}
Some users mentioned that they had wanted a security diagnosis service such as this to be available~(code:service-meets-needs-of-diagnosis, n=370), for example, \textit{``Since security software on PCs and smartphones alone cannot check for router vulnerabilities, I feel that this is a valuable means of checking.''} This further implies that a service such as this was filling a perceived gap.

\textbf{Prompting re-diagnosis~(Usefulness)}
We also found that the follow-up campaign to prompt re-diagnosis was useful for users~(code:notification-of-follow-up-campaign, n=202). For example, 
\textit{``I had forgotten that I had even used the service myself, but I received an email invitation to inspect, which prompted me to have a re-diagnosis and gave me reassurance. I thought it would be good to receive periodic push-type diagnosis recommendations. ''}
As discussed above, most users were willing to use the service again, but only 11\% of those users actually did so.
The implication is that users may intend to take security measures but tend to forget; thus, periodic prompts are necessary.

\textbf{Improving security awareness~(Usefulness)}
In some cases, our service contributes to improving users' security awareness~(code:improvement-of-security-awareness, n=48). For example, \textit{``I had no idea to defend even my home router. Having got that knowledge is a big step forward.''}

\textbf{Easy to use~(Usability)}
Most users mentioned that the procedure to take a diagnosis is simple and easy~(code:ease-of-diagnosis, n=1628). For example, a user said about easy operations to diagnose, \textit{``It is easy to diagnose without any difficult operations. [..]''}
Another user mentioned that users do not need security knowledge, \textit{``Security checks can be done without requiring any specialized knowledge.''}

\textbf{Short diagnosis time~(Usability)}
In addition, most users felt that the short diagnosis time was a good point~(code:short-diagnosis-time, n=904). 
Specifically, most users were satisfied with the short time between requesting a diagnosis and receiving the results, for example, \textit{``It was good to see the results immediately.''}

\textbf{Trustworthy site operator~(Trustworthiness)}
Users also mentioned that the site operator was trustworthy because it is not a for-profit organization but a public university~(code:good-trustworthiness-of-site-operator, n=156). For example, \textit{``I decided to use this service again because I had a follow-up to a previous inspection. Because it is a service provided by a university, not a company.''}  Another user said, \textit{``It is easy to use, the results come out immediately, and it is trustworthy because it has been featured in the media and is academic.''}
In addition, some users mentioned that the absence of advertisements on the website is a good point. It means that the non-profit operator and non-profit service would be a factor in gaining the trust of users.

\subsection{Representative responses on bad points}
\label{sec:response-bad-points}
\textbf{Summary}
The right part of Figure~\ref{fig:good-bad-points} shows the coding results of bad points.
Most users~(n=2471) mentioned that there was no bad point in particular.
This could be read as indicating that the implementation of the service meets users' expectations. However, to caveat this seeming success, some users indicated wanting to know the details of the diagnosis~(n=461). 
Some users indicated a lack of trustworthiness of email communication~(n=325).
Moreover, users want regular reminders to encourage diagnosis~(n=118). Some technical issues did arise, as some users encountered technical troubles due to the Cookie settings of their browsers~(n=82).

\textbf{Lack of diagnosis details~(Usefulness)}
Users would like more details about the diagnosis~(code:lack-of-details-of-diagnosis, n=461). For example, \textit{``The results showed no problems, but that alone made me wonder, "Is this true?" I felt that it would be more convincing if the details of the judgment leading up to the "no problem" result were also displayed.''}
When designing the service, we were mindful to address the detail of terminology to different levels of expertise, focusing on non-experts and simplicity. As we discussed above, this design principle contributes to ``easy-to-use,'' but it is insufficient for the skilled user and leads to dissatisfaction. Established design heuristics \cite{nielsen1994enhancing} indicate that expert users could be served with `shortcuts' to advanced features, but that not having these is a minor design issue (as it does not necessarily block use of the service).

\textbf{No regular reminders to encourage diagnosis~(Usability)}
Users wanted regular reminders to encourage diagnosis~(no-regular-follow-up-campaign-email, n=118). For example, \textit{``It would be appreciated if you could notify me by e-mail or other means for diagnosis on a regular basis.''} 
In addition, regular reminders would improve the trustworthiness of emails from the service; for example, \textit{``I hadn't received an email in a while and didn't remember the service. I was about to report it as spam. Please include an email once a month or so to remind me. [..]'' }
Similar to the analysis of good points, the analysis of bad points also indicates that a regular reminder needs to be performed. Some consideration should be given to the expected regularity of behaviours (as an aspect of behavior change design \cite{fogg2010behavior}).

\textbf{Cookie and captcha issues~(Usability)}
Most complaints about usability issues came from system specifications. Specifically, users had trouble due to the Cookie settings of their browsers~(code:troubles-due-to-cookie, n=82). For example, \textit{``I could not see the diagnostic results because my browser was in private mode, so I had to start over. I wish there had been a notice beforehand.''}
In addition, some users complained about i-am-not-a-robot confirmation~(code:bad-user-experience-with-recaptcha, n=41), for example, \textit{``The photo selection screen for i-am-not-a-robot confirmation is small and difficult for the elderly to see. We are often asked to redo the process. Please take this into consideration.''} 
For usability, these barriers must be removed.

\textbf{Suspicious email about follow-up campaign ~(Trustworthiness)}
Several users were aware of the growing phishing attacks in recent years and were concerned about the authenticity of our emails and websites. Specifically, some users complained about the email with a link because it seemed like a phishing email~(poor-trustworthiness-of-email-communication, n=325). For example, \textit{``This is not necessarily a bad thing, but with the prevalence of phishing e-mails, I was often hesitant to click on the link in the e-mail that prompted me to retest. I would feel more comfortable if there was some mechanism to assure me that the e-mail I received was not a phishing e-mail. ''}
There are technical mechanisms to help ensure the authenticity of emails, such as DMARC (Domain-based Message Authentication Reporting and Conformance), but it is difficult for users to understand. Maybe more trustworthy channels, such as a dedicated smartphone app, would be better. Studies of ISP support for customers have found that when there are notifications of a malware infection on a home network, customers may regard the notification as being spam~(e.g.,~\cite{rodriguez2022difficult}).

\textbf{Trustworthiness of web site~(Trustworthiness)}
Users were also suspicious that our website might be a fake site~(code:poor-or-unclear-trustworthiness-of-site-itself, n=138). For example, \textit{``I had a momentary concern when performing the diagnosis that this service itself might be a scam site. This may be something that requires the cooperation of the government (approval, certification, etc.), but I thought it would be better if there was something more to prove the security of the service.''}
Similar to emails, explaining the authenticity of the website to users, especially those with limited skills, is an unresolved issue.

\textbf{Lack of name recognition~(Trustworthiness)}
Some users thought that the service should have more name recognition for trustworthiness~(code:low-name-recognition, n=45). For example, \textit{``I didn't know about this site at all until I found from the news, so I was a little worried when I used it for the first time. I want more people to know about this so that people can use the Internet securely.''}
Behavior change approaches within cybersecurity have emphasized the importance of the entity which acts as the `messenger' delivering information to users \cite{coventry2014scene}, where here we have seen that this has affected the perception of some service users.
To increase the name recognition and trustworthiness of our service, more promotion would be required, such as cooperation with the government.

\subsection{Responses of users rating the service as not helpful and users with security issues}
\label{sec:user-rating-bad-or-with-security-issue}

\textbf{Users who rated the service as not helpful.}
Here, we focus on users who answered that the service was not helpful~(n=31). Surprisingly, some users rated the service as unhelpful but did not point out any bad points. Specifically, 19 out of 31 users answered that there were no bad points.
Similarly, there was no difference in their responses for the good points, for example,  ease-of-diagnosis~(n=12), reassurance-after-diagnosis~(n=9), and short-diagnosis-time~(n=4).

\textbf{Users with security issues.}
Next, we focus on answers from users with security issues~(n=3). The results showed that there was no difference in the good points: free-of-charge~(n=2), improvement-of-security-awareness~(n=1), good-trustworthiness-of-site-operator~(n=1), and ease-of-diagnosis~(n=1).
As well as the good points, there were no differences for bad points, as all responses were coded as cannot-name-specific-bad-points~(n=3).

\section{Results: notification to users with security issues}
\label{sec:notification-using-amii}

In this section we answer RQ2, discussing the identified risks and remediation rate.

\subsection{Identified risks}
\label{subsec:identified-risks}

Of the \updatenumC{98,722}{114,747} users, \updatenumY{524~(0.53\%)}{585~(0.51\%)} users had been identified that their devices had security issues. 
We identified \updatenumC{138~(0.14\%)}{171~(0.15\%)} users whose devices had been detected as malware-infected, and \updatenumC{389~(0.39\%)}{417~(0.36\%)} users whose devices had been detected as vulnerable.
\update{Three users had both malware infections and vulnerabilities.}
In cases of \updatenumY{379~(72\%)}{432~(74\%)} users with risks, one risk was detected at a diagnosis, and in the remaining \updatenumY{28\%}{26\%} cases multiple risks were identified. In the largest case, \updatenumY{4}{4} risks were found at the same time. 

\update{Overall,  malware infection~(171 cases),  known default ID~(154 cases) ,  risky protocol (Telnet)~(121 cases) , and  old firmware~(113 cases), were the most frequently detected.
As for the categories of IoT devices with security issues, we identified routers, webcams, NAS, and firewalls.
Detailed numbers on each risk and IoT category is provided in Appendix~\ref{appendix:details-of-notification-results}.}

\subsection{Remediation rate}
\label{subsec:recheck_rate_and_remediation_rate}

\begin{table}[t]
    \centering
    \caption{\update{Notification results}}
     \label{tab:notification-stats}
    \footnotesize
    \begin{tabular}{l@{\hskip1pt}r@{\hskip4pt}r@{\hskip5pt}r}
    \\\toprule
     & Malware & Vul.(all) & Vul.{\scriptsize (exc. def ID/PW$\dag$)}
    \\ \midrule
         Users of our service &  \multicolumn{3}{c}{114,747} \\ \hdashline
         Users with issues &  171 & 417 & \todo{311} \\
         Users who did re-diagnosis & \todo{67}$\ddag$ & \todo{151} & \update{117}\\
         Users with remediation & \todo{59} & \todo{75} & \update{59}\\
    \bottomrule
    \end{tabular}
    \scriptsize
    \vspace{-2mm}
    \flushleft{
    $\dag$~Except known default ID, known default credentials, and weak default Wi-Fi pass.
    \\
    $\ddag$~Users who re-diagnosed at least 24 hours after the malware infection was detected.
    }
\end{table}

To calculate the re-diagnosis rate and remediation rate, we took into account diagnoses in multiple environments. The users may diagnose the security risks in several places, for example, at home, and at the workplace. In such cases, the re-diagnosis rate and remediation rate must be calculated for the environment where the risks are identified. To cope with the cases, we check the autonomous system~(AS) number of the user's IP address.

Here, we focus on the responses of users who were notified of malware infections and vulnerabilities. Of the \updatenumC{138}{171} users who were notified of malware infections, \updatenumC{84~(61\%)}{\todo{108~(63\%)}} performed re-diagnosis, and \updatenumC{49~(36\%)}{\todo{67~(39\%)}} performed re-diagnosis after 24 hours or more from the time the risk was detected.
The need for a certain period of time is due to the mechanism of malware infection detection, as described in Section~\ref{subsec:method-amii}.
\update{Of the \updatenumC{49}{\todo{67}} users who re-diagnosed at least 24 hours after the malware infection was detected, \updatenumC{44~(90\%)}{\todo{59~(88\%)}} were judged to be in a clean state, and remediation was confirmed.}

For users whose vulnerabilities were detected, \updatenumC{140 of 389 users~(36\%)}{\todo{151} of 417 users~(36\%)} re-diagnosed. The re-diagnosis rate for users whose problems were detected was generally higher than the overall re-diagnosis rate~(\updatenumC{28\%}{29\%}) shown in Section~\ref{subsec:usage_trends}.
\update{Of the \updatenumC{140}{\todo{151}} users who re-diagnosed after the vulnerability was detected, \updatenumC{69~(49\%)}{\todo{75~(50\%)}} showed remediation.}

Among the risks shown in Table~\ref{tab:diagnositc_items_and_recommended_actions}, Known default ID, Known default credentials, and Weak default Wi-Fi passwords continue to be detected even after the password are changed since we did not check the password actually set on the device. If the devices became not visible from the Internet, the risks were counted as remediated.
Of the \updatenumC{389}{417} users with risks of vulnerabilities, \updatenumC{291}{\todo{311}} users were detected with risks other than those listed above. Of these, \updatenumC{110~(38\%)}{\todo{117~(38\%)}} re-diagnosed, and \updatenumC{57~(52\%)}{\todo{59~(50\%)}} showed remediation of them.

Finally, we investigate remediation rate of users with multiple risks.
As discussed in Section~\ref{subsec:identified-risks}, we diagnosed 10 types of risks. As a result, 
\updatenumY{146~(28\%)}{151~(26\%)}
users had multiple risks.
Of the users with multiple risks, we identified \updatenumY{11~(8\%)}{12~(8\%)} of users had multiple devices.
Of the users with multiple risks identified, \updatenumY{37~(25\%)}{38~(25\%)} users remediated at least one risk, \updatenumY{32~(22\%)}{33~(22\%)} of users remediated all risks.

\begin{figure}[tb]
    \centering
    \includegraphics[width=\linewidth, trim={4.5mm 5mm 2mm 0},clip]{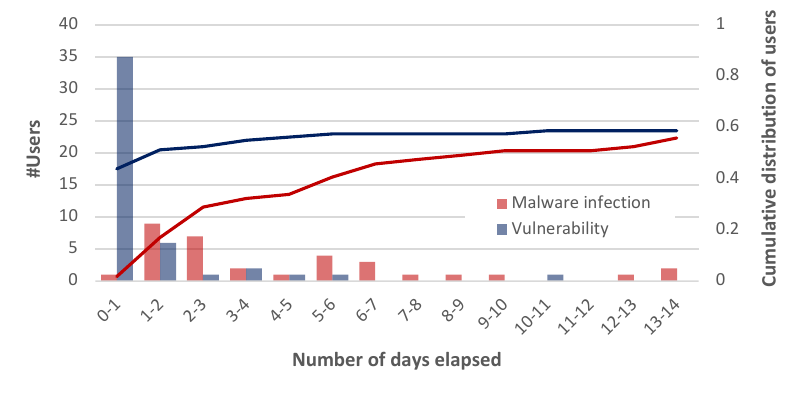}
    \caption{Time between notification and remediation confirmed~\updatefigC
    }
    \label{fig:required_time_until_recheck}
\end{figure}

Figure~\ref{fig:required_time_until_recheck} shows the time required for each user to re-diagnose after a malware infection or vulnerability is detected.
Here, remediation for vulnerability means that at least one vulnerability has been addressed in case a user had multiple vulnerabilities.
In both groups of malware infection and vulnerability, the majority of users who performed re-diagnosis performed it within one day of the diagnosis date. 

\section{Results: survey on taking measures}
\label{sec:survey-on-taking-measures}
Here we focus on RQ3, and users' obstacles to remediation. Table\
\ref{tab:overview_of_group_assignment_and_respondents} shows the results of questionnaires to users with remediation confirmed and the users with remediation not confirmed. 
Since the mail delivery rate was 96\% due to nonexistent or invalid addresses, the number of recipients got slightly lower than the number of survey targets, but still
we received responses from a large number of users.

\begin{table}[t]
    \centering
    \caption{Subjects of questionnaires and answers}
    \label{tab:overview_of_group_assignment_and_respondents}
    \footnotesize
    \begin{tabular}{l@{\hskip2pt}cccr@{\hskip7pt}r@{\hskip3pt}}
        \\\toprule
         & \multicolumn{2}{c}{\shortstack{Remediation confirmed}} & \multicolumn{2}{c}{\shortstack{Not confirmed}}&\multirow{2}{*}{Total}\\
        \cmidrule(lr){2-3}
        \cmidrule(lr){4-5}
        & Malware inf.* & Vul. & Malware inf. & Vul. &\\
        \midrule
        \#Subjects &26&41&58&242&367\\
        \#Answers &11
        &17
        &13&49&90\\
        \bottomrule
    \end{tabular}
    \scriptsize
    \vspace{-2mm}
    \flushleft{* There would be false negatives caused by our detection method.}
\end{table}

\textbf {Users with remediation confirmed.}
First, we analyze the users whose devices were remediated, and the remediation was confirmed by re-diagnosis.

Of the 11 users who had detected malware infection, 9 (82\%) 
responded that they had taken measures. Of these, 6  users selected ``measures we proposed'' (firmware updates and device reboots), 2 users selected ``we prepared a replacement device.'' 1 user selected ``Other'' and reported ``changing the name of the default Wi-Fi setting,'' which is not a reasonable measure because changing the SSID name would not remediate the malware infection. It is assumed that the remediation was caused by a reboot when the SSID name was changed.

Of the 17 users who had vulnerabilities detected, 16 (94\%) responded that they had taken measures. Of the 16 users, 
9 responded that they had taken the measures we proposed, 6 answered that they had prepared replacement devices. 1 chose ``Other'', reporting that another device with a similar model number had been detected as a false positive.

The above result suggests that the majority of the users with remediation confirmed by the re-diagnosis had implemented appropriate measures. It was clear that notifications of our service contributed to the remediation. 
The users who improved without taking any measures seemed to be due to spontaneous recovery or false negatives of the detection method.

\textbf{Users with remediation not confirmed.}
\label{subsec:results_of_those_who_still_have_issues} 
Here, we focus on the users whose remediation is not confirmed, specifically, users who did not re-diagnose and users who did re-diagnose but whose risks still remained.
\begin{table}[t]
    \centering
    \caption{Answers from users with remediation not confirmed. }
    \label{tab:survey_result_of_those_who_still_have_problems}
    \footnotesize

    \begin{tabular}{@{\hskip2pt}p{3.0cm}@{\hskip3pt}l@{\hskip1pt}l@{\hskip3pt}l@{\hskip3pt}l@{\hskip3pt}l@{\hskip2pt}}
        \\\toprule
        & \multirow{3}{*}{\shortstack{Malware \\infect.}} & \multicolumn{4}{c}{Vulnerability} \\ 
        \cmidrule(lr){3-6}
        & & \multirow{2}{*}{\phantom{00}Total} & \multicolumn{3}{c}{\#Vulnerabilities} \\
        \cmidrule(lr){4-6}
        & & & 1 & 2 & 3 \\
        \midrule
        Users & 13 & 49 & 32 & 12 & 5\\
        \cmidrule(lr){1-6}
        Tried to take measures & 6 (46\%) & 29 (59\%) & 20 (63\%) & 6 (50\%) & 3 (60\%) \\
        Identified devices with risk & 5 (38\%) & 27 (55\%) & 18 (56\%) & 6 (50\%) & 3 (60\%)\\
        Completed the measures & 4 (31\%) & 25 (51\%) & 17 (53\%) & 5 (42\%) & 3 (60\%) \\
        \bottomrule
    \end{tabular}
\end{table}
\begin{table}[t]
    \centering
    \caption{Why the users did not take measures.}
    \label{tab:reason_why_not_trying_remediation}
    \footnotesize
    \begin{tabular}{@{\hskip3pt}l@{\hskip4pt}l@{\hskip4pt}l@{\hskip3pt}}
        \\\toprule
        Reason (Multiple choices allowed) & Malware & Vul. \\ 
        \midrule
        Did not feel the need for remediation & 1~(14\%) & 0~(0\%)\\
        Did not know how to take measures  & 2~(29\%) & 6~(30\%) \\
        Taking measures is technically difficult  & 2~(29\%)  & 5~(25\%) \\
        Forgot to take measures   & 1~(14\%) & 0~(0\%)\\
        Measures would affect the usage of the device   & -  & 4~(20\%)\\
        Replacement of the device causes financial burden   & 1~(14\%)  & 6~(30\%)\\
        Not owner of the device* & - & 4~(20\%)\\
        False positives* & - & 2~(10\%)\\
        Remediation confirmed using another email address*   & 1~(14 \%) & -\\
        \midrule
        Users who answered this questionnaire item & 7 & 20\\
        \bottomrule
    \end{tabular}
    \scriptsize
    \vspace{-2mm}
    \flushleft{* Codes generated from responses of ``Other,'' not set as pre-defined choices.}
\end{table}
\update{Table}~\ref{tab:survey_result_of_those_who_still_have_problems} shows answers from uses with remediation not confirmed. Seven of the 13 users (54\%) of users with malware infections answered that they did not attempt any measures. Table~\ref{tab:reason_why_not_trying_remediation} shows the reasons. Only one user answered, ``did not feel the need for remediation,'' and almost all users understood the need for remediation. However, three users answered either ``did not know how to take measures'' or ``taking measures is technically difficult'' (one user answered both). As for 2 open-ended responses about the content of ``Other,'' one was coded as ``replacement of the device causes financial burden.'' The other was that the user checked with a different e-mail address after having taken measures and confirmed the remediation.

Of the six users (46\%) who indicated that they had attempted to take action, five had successfully identified the device, and four of them had completed the action. Of these four, one had not been re-diagnosed at the time of the survey, and the other three had been re-diagnosed, but the time was less than 24 hours from the diagnosis. Two of the four users had a re-diagnosis after completing the questionnaire and remediation. Users who gave up on implementing measures after the devices were identified responded, ``It was difficult to operate the devices for firmware updates and reboots.''

Of the 49 respondents from users whose vulnerabilities were detected, 20 (41\%) indicated that they did not try to take any measures. 
No user answered ``did not feel the need for remediation''. Therefore, all users understood the need for remediation, but eight users answered ``did not know how to take measures'' or ``taking measures is technically difficult'' (three users answered both). In addition, six users~(5 out of 6 were users with end-of-life devices) answered that they ``replacement of the device causes financial burden,'' indicating that the financial burden was an obstacle even if the user had the will to take measures. 
As for 6 open-ended responses from users who chose ``Other,''
4 were coded as ``Not owner of the device'' and 2 were coded as ``False positives.'' 

Of the 29 users (59\%) who indicated that they had attempted to take measures, 27 successfully identified the device, and 25 of these users indicated that they were able to take measures. Of these 25, 16~(64\%) had not been re-diagnosed at the time of the survey. Of these 16, one user was re-diagnosed after completing the questionnaire, and remediation was confirmed. 
Of the two users who gave up in the process after the devices were identified, one user reported ``difficulty in operating the devices (changing authentication settings, updating firmware, stopping Telnet, etc.).''

Approximately half of the users with unconfirmed remediation attempts took measures, and most of them completed the measures. Surprisingly, many of them did not re-diagnose their IoT devices, and their security situations after taking measures were not clear. 
Users may not have performed a re-diagnosis because of their trust in our service. Specifically, users may have trusted that recommended measures make the devices a safe state.  Here, the trust in the service would be a double-edged sword; if users trust the service enough, the users follow the instructions, but users may not re-diagnose. That can be regarded as a delegation of responsibility~\cite{Dourish2004}, but prior research has shown that user action may not achieve the result they think it does \cite{TheThingDoesntHaveAName}.

\textbf{Analysis of motivation and ability based on questionnaire answers.}
According to the B=MAP model~\cite{Fogg-Behavior-Model}, prompted users who did not take measures lacked motivation and/or ability. 
According to Table~\ref{tab:reason_why_not_trying_remediation}, both users with malware-infected devices and vulnerable devices had motivation. As for the ability, approximately half of the users had issues in cases of both malware infections and vulnerabilities.

\section{Discussion}

Here we reflect on the lessons learned from the \updatenumC{16}{27}-month operation of the service, supported by the findings from the user scans and survey responses.

\label{sec:lessons}
\textbf{What are the key success factors of user engagement of the security diagnosis service?}
Promotion is important because users need to visit the website proactively. As discussed in Section~\ref{sec:service-deployment}, promotion in the TV program and promotion through reminder emails to recommend re-diagnosis were effective. In terms of service characteristics, as discussed in Section~\ref{sec:user-survey-procedure}, it is important that the diagnosis be free of charge so that users who visit the site will not hesitate to use it, that the operation be simple so that they can easily undergo the diagnosis, and that the results be immediately available.

\textbf{Is there a gap between users' expectations and the value of the services?}
For most users, the answer is no. According to the questionnaire, 96\% of the users positively rated the service. Additionally, most users answered that the service had no bad points. As for the negative points, users' feedback is mostly on technical properties such as the use of a cookie and the CAPTCHA, and not on the diagnosis itself.
Given the large number of respondents who said they were glad to have reassurance, it seems that users utilized our service with the expectation of receiving reassurance and were satisfied with the results, as expected.

\textbf{False reassurance.}
\update{The diagnostic results may include false negatives, which could lead users to gain false reassurance. We have made efforts to reduce false reassurance by informing users of the limited diagnostic capabilities; the service's top page states that diagnostic accuracy is not perfect and there may be false negatives. 
Additionally, the result page states  ``within the scope of this service.''
}

\update{False reassurance also arises from users' misunderstandings on the diagnosis scope.}
We found that a few users did not understand the scope of the diagnosis and had a false sense of reassurance~(n=8). For example, our service only diagnosed malware infection and vulnerability of IoT devices, but the following user felt reassurance.
Specifically, a user said \textit{``I was worried because of the frequent reports of cyber attacks and credit card fraud, but I did not know how to look into it, so I was relieved for the time being.''}
\textcolor{black}{Every security diagnostic service has this side effect. This limitation could be acceptable if the benefits of the service outweigh the drawbacks.}

\color{black}
\textbf{Who benefits from the service?}
100k users used the service and benefited from the diagnosis results.
For the majority of users, who were already motivated enough to visit the website, they received a clean scan which provided reassurance.
Only a small number of users were left with lingering questions about whether their devices are secure.

The B=MAP behavior change model~\cite{Fogg-Behavior-Model} can be used to articulate this, wherein successful enactment of a new behavior relies on a combination of  
sufficient Motivation and Ability, and a Prompt that helps a person reach that target behavior from where they are currently. 
The promotion of the service on TV (Table~\ref{tab:timeline}) would have served as a `signal' prompt to check the security of potentially vulnerable devices; the service itself is a `facilitator' prompt which aims to check device security for the user (rather than them doing this themselves). Users with high Motivation and high or low Ability may react to the TV prompt, by following the links to the service.

User experience would differ once using the service, as to whether they need to take action after the scan, and whether they have relatively high or low Ability - users would need to understand and accept the scan results (clean or not), or if action is needed, be able and confident enough to take the advised action. 
The service is assumed to only reach those who monitor the media channels where the service was promoted, have sufficient Ability to recognize content in the service (and risk) description that relates to them, and have Motivation enough to be willing to check their devices (even if they believe their devices are not at risk -- hence the service being free to use, not requiring further `Ability' to pay for it). As such, low Motivation users who care less about device security are much less likely to respond, which is a limitation that then influences the range of users.

\color{black}

\subsection{Limitations}
\textbf{Self-selection bias.} One limitation in our work is that there is a likely self-selection bias among the respondents, of people who are proactive about security, responding to the `prompt' to take action and conduct a scan.
\textcolor{black}{
A lot of research on real-world security solutions studies users who choose to use the solutions, such as a network scan tool~(a part of antivirus)~\cite{AllThingsConsidered}, browser security extensions~\cite{privacy-preserving-browser-extension}, and security keys~\cite{security-key}. By definition, any research on real-world service has selection bias. Even with the bias, the major contributions of our study -- helping approximately \update{115k} actual users and understanding their experience -- are worthwhile.}

\textbf{Accuracy and coverage of diagnosis.}
Another limitation is the potential for false positives and negatives: as discussed in Section~\ref{subsec:system-design}, our detection method has the potential to cause false positives and negatives. \textcolor{black}{In addition, coverage of IoT devices is limited.} However, the detection method is relying on current technical solutions, with limitations inherent to most if not all, scanning tools for consumer IoT devices.

\update{
\textbf{Limitation due to external monitoring.}
As discussed in Section~\ref{subsec:system-design}, the scope of the service is limited to vulnerabilities that can be diagnosed from the Internet. 
}

\update{
\textbf{Survey responses from users with problems.}
Regarding the survey on the good and bad points of our service mentioned in Section~\ref{sec:user-rating-bad-or-with-security-issue}, there were three responses from users who experienced issues. More responses are needed to generalize the findings. Also, the survey responses are provided by likely non-expert users, self-assessing their understanding and enactment of remediation advice (where prior work has shown that users may struggle to understand complex IoT instructions \cite{TheThingDoesntHaveAName}). As the service continues, we expect to be able to associate return scans with survey responses.  
}

\subsection{Recommendations}
Here, we describe recommendations obtained from the deployment of our service.

\textbf{Lower the cost of entry and usage.} This includes reducing the effort of finding the service, as was done here through news promotion and reminders to recommend re-diagnosis. Also, simplicity lowers the barrier to using the service. 

\textbf{Respond to needs for assurance.} Consider that even users without problems may want -- and value -- assurance that their devices have no security issues.  
From our results, participants welcomed confirmation that their devices had no detectable security issues. 

\textbf{Balance comprehensibility with expectation-setting.} Users may have overestimated what the service can do. There is a trade-off between providing a precise explanation of what the scan can and cannot do, and providing a short and simple explanation that users can actually comprehend. Any attempt to be short and simple risks that some users will make excessive assumptions about the capability of the service. According to the questionnaire answers, some users wanted detailed information about the diagnosis, so shortcuts/accelerators can help \cite{nielsen1994enhancing}. A similar consideration about balancing generic and specific advice is discussed in the provision of preventative measures \cite{reeder2017152}, where here we find lessons are transferable also to detection measures.

\textbf{Provide customized support.} When serving \update{115k} users, how feasible is it to handle user requests for support? We provided users with an email option and a button on the results page. 
During the operation of the service, we received \updatenumC{eight}{eight} requests~(\updatenumC{six}{four} emails and \updatenumC{two}{four} requests from the diagnosis result page) asking for direct support from users with security issues, which is a tiny fraction of the \updatenumC{524}{585} users with security issues. This suggests that even large-scale services might offer direct support channels without having to staff costly support infrastructure.
Other studies~\cite{rodriguez2022difficult} also found that only small numbers of users need targeted support and that users can often remediate issues themselves.

\section{Related work}

\textbf{Notification campaigns.}
Many notification campaigns have been studied for their effectiveness in remediation~\cite{maass2021effective, Comparing-Large-Scale-Notifications, Best-Practices-for-Notification}
As for the notifications on consumer IoT devices, Cetin et al. conducted a notification campaign for devices infected by Mirai malware, \update{via a partner ISP}~\cite{ccetin2019cleaning}. Notification with a walled garden, which quarantines the infected devices, achieved a high remediation rate. In contrast, they found that email-only notifications have no observable impact.
\update{As notifications via ISP, the Japanese government, a national research institute, and ISPs are collaborating to investigate devices that may be misused in cyber attacks and to issue notifications to the owners of such devices~\cite{notice}.}
The WarpDrive project~\cite{warpdrive} has proposed a method using a dedicated application such as smartphone apps and browser extensions. 
This method enables long-term monitoring and notification of each user, but on the other hand, the installation of the dedicated application is burdensome for the users.

\update{
Different from the above studies, we have shown that Web services that do not depend on a specific ISP or require the installation of an agent can support remediation.
}

\textbf{User studies.}
Users' reactions to the security notifications have been surveyed in many studies.
Bouwmeester et al. conducted a survey of 17 users who had infected devices and identified only four users were able to complete all remediation steps~\cite{TheThingDoesntHaveAName}. 
Rodriguez et al. have conducted a survey of users whose IoT devices were infected by QSnatch which is a persistent IoT malware targeting network-attached storage~(NAS)~\cite{rodriguez2022difficult}. They identified that uses remediated right or closely after receiving a notification.

Studies on user support and assistance have been conducted.
As a security helper about privacy concerns, IoT Privacy Assistant~\cite{das2018personalized,colnago2020informing} to help discover \update{nearby} IoT resources and configuration of \update{data sharing} settings has been proposed. 
Information security was analyzed in analogous to healthcare problems, and the concept of information healthcare has been proposed~\cite{flechais2015towards}.
Poole et al. surveyed the informal technical support of computers~\cite{PooleCMGE09}. The authors identified that family and friends of interviewees often did not know where to look for helpful resources and whom to contact for problems. 
Nicholson et al. evaluated community-driven initiative~(CyberGuardians) for older adults~\cite{CybersecurityGuardians}. 
Havron et al. proposed clinical computer security to help victims of intimate partner violence, e.g., tracking using spyware~\cite{ClinicalComputerSecurity}.

\update{As a kind of user support, security topics have been discussed in a Q\&A forum~\cite{hasegawa2022understanding}. This included security questions about cyberattacks, authentication, and security software. It was found that users had difficulty explaining their issues. 
A similar study on Twitter was conducted by Pattnaik et al.~\cite{Pattnaik_2023}.}

\update{In contrast to these user studies, we focused on what users expect from Web-based security diagnostic services, and clarified what they consider good and bad.}

\section{Conclusion}
We designed and operated the web-based security diagnosis service for IoT devices. During the \updatenumC{16}{27}  months of operation, we gained \updatenumC{98,722}{114,747} users, identified \updatenumC{524}{585} users whose devices have security issues, and confirmed \todo{that 133 users showed improvement}. 
We conducted a user survey on our service and received responses from 4,103 users. Through the analysis of the responses, we identified that many users saw the value of the service in the reassurance they received, even when no problems were found.
We conducted a survey of users having security issues and found that most users were willing to take measures, but a lack of security ability was a major obstacle.

\section*{Acknowledgements}

This research was partly supported by the Dutch Research Council (NWO) via the RAPID (Grant No. CS.007) and INTERSCT (Grant nr. NWA.1160.18.301) projects.
A part of these research results were obtained from the commissioned research~(No.JPJ012368C08101) by National Institute of Information and Communications Technology (NICT) , Japan. 
This work was supported by JSPS KAKENHI Grant Numbers 23K11099 and 21KK0178.

We obtained darknet data through the NICT CYNEX/Conexus-S collaboration framework and used the data to detect malware infections. We also used 00One's ``Karma'' IoT scanning engine to detect security issues.

\bibliographystyle{plain}
\bibliography{reference}

\appendix

\section{Notification and recommended measures (Translated from Japanese)}
\label{sec:notification-messages}

For users with security issues, our service displayed a security notification and recommended the following measures. 
These are translated from Japanese, and two researchers ensured the quality of the translation.

\textbf{Malware infection.}
Your device (router, etc.) is communicating suspiciously and may be infected with malware.
Reboot the device. Then, update the firmware according to the device manual. Many devices have manuals available on the Internet. To find the manual, use a search engine such as Google and search for the manufacturer name, device model, and ``manual'' as keywords. Due to the diagnostic mechanism, the diagnosis cannot be performed correctly right after the measures have been taken. Be sure to wait at least 24 hours before re-diagnosis.

\textbf{Risky protocol (Telnet).}
An old communication program (Telnet) is running. There is a risk of unauthorized access or malware infection. Immediate countermeasures are required. 
Stop Telnet according to the device manual. Many devices have manuals available on the Internet. To find the manual, use a search engine such as Google and search for the manufacturer name, device model, and ``manual'' as keywords.

\textbf{End of support.}
A device that the manufacturer no longer supports has been detected. 
Even if vulnerabilities are discovered in the devices, the vulnerabilities are not fixed. Thus, there is a risk of unauthorized access or malware infection. Immediate countermeasures are required. 
It is difficult to continue using the device safely. Please consider replacing it with a new device.

\textbf{Admin password not set.}
A device that is being used with an unset password has been detected. There is a risk of unauthorized access or malware infection. Immediate countermeasures are required.
Set the password according to the device manual. Many devices have manuals available on the Internet. To find the manual, use a search engine such as Google and search for the manufacturer name, device model, and ``manual'' as keywords.

\textbf{Known vulnerability.}
A vulnerable device has been detected. Immediate countermeasures are required.
Follow the manual of the device to update the firmware. Many devices have manuals available on the Internet. To find the manual, use a search engine such as Google and search for the manufacturer name, device model, and ``manual'' as keywords.

\textbf{Old firmware.}
A device with firmware that is not up-to-date was detected. Vulnerabilities in the devices have not been corrected and may result in unauthorized access or malware infection. Immediate countermeasures are required. 
Follow the manual of the device to update the firmware. Many devices have manuals available on the Internet. To find the manual, use a search engine such as Google and search for the manufacturer name, device model, and ``manual'' as keywords.

\textbf{Known ID.}
A device with a publicly available initial ID has been detected. If easily guessable passwords are used, there is a risk of unauthorized access or malware infection. Please note that we do not inspect the IDs and passwords actually set on the devices. Therefore, even if the ID/password is changed, the inspection result will not change.
Please follow the manual of the device to change the password. Many devices have manuals available on the Internet. To find the manual, use a search engine such as Google and search for the manufacturer name, device model, and ``manual'' as keywords. To make passwords hard to guess, choose a combination of upper and lower case letters, numbers, and symbols, and choose as many random ones as possible rather than meaningful English words. In general, the longer the password, the harder it is to guess. Also, do not use the same password that you use elsewhere.

\textbf{Known credential.}
A device with a publicly disclosed initial ID and password has been detected. 
If the configuration has not been changed,
there is a risk of unauthorized access or malware infection. Please note that we do not inspect the IDs and passwords actually set on the devices. Therefore, even if the ID/password is changed, the inspection result will not change.
Please follow the manual of the device to change the ID/password.
Many devices have manuals available on the Internet. To find the manual, use a search engine such as Google and search for the manufacturer name, device model, and ``manual'' as keywords. To make passwords hard to guess, choose a combination of upper and lower case letters, numbers, and symbols, and choose as many random ones as possible rather than meaningful English words. In general, the longer the password, the harder it is to guess. Also, do not use the same password that you use elsewhere.

\textbf{Vulnerable default Wi-Fi password.}
A Wi-Fi device whose initial password is easy to guess has been detected. Someone may access the Wi-Fi. Please note that we do not inspect the passwords actually set on the devices. Therefore, even if the password is changed, the inspection result will not change.
Please follow the manual of the device to change the Wi-Fi password~(WPA key). Many devices have manuals available on the Internet. To find the manual, use a search engine such as Google and search for the manufacturer name, device model, and ``manual'' as keywords. To make passwords hard to guess, choose a combination of upper and lower case letters, numbers, and symbols, and choose as many random ones as possible rather than meaningful English words. In general, the longer the password, the harder it is to guess. Also, do not use the same password that you use elsewhere.

\textbf{No authentication.}
A device that can be operated without ID/password authentication has been detected. There is a possibility of being operated from outside.
If you are using the equipment for applications that require security, please consider replacing it with a new device.

\section{Contents of FAQ (Translated from Japanese)}
\label{sec:faq-details}

\update{Table~\ref{tab:faq-contents} shows FAQ contents translated from Japanese.}

\begin{table}[h]
    \centering
    \caption{Contents of FAQ}
    \label{tab:faq-contents}
    \footnotesize
    \begin{tabular}{p{8cm}}
    \toprule
    What is Malware? What does it mean to be infected with malware?
    \\How do \amii~detect malware?
    \\ What should I do if my device is infected with malware?
    \\ What should I do if the test result does not change even after restarting the router?
    \\ What types of security issues in routers can be detected?
    \\ How do \amii~detect security issues?
    \\ What is firmware? How should I update the firmware?
    \\ What are vulnerabilities?
    \\ Are computers and smartphones subject to inspection?
    \\ What is an email address used for?
    \\ What information do \amii~collect and record?
    \\ How is the collected information handled?
    \\ How can I find the manual for the device?
    \\ What is Telnet?
    \\ What is the initial ID and password?
    \\ How to choose a hard-to-guess password?
    \\ Are there any false positives/negatives in the diagnostic results?
    \\ \bottomrule
    \end{tabular}

\end{table}

\section{Questionnaire about engagement}
\label{sec:questionnaire-about-engagement}
\begin{figure}[t]
    \centering
    \includegraphics[width=0.9\linewidth]{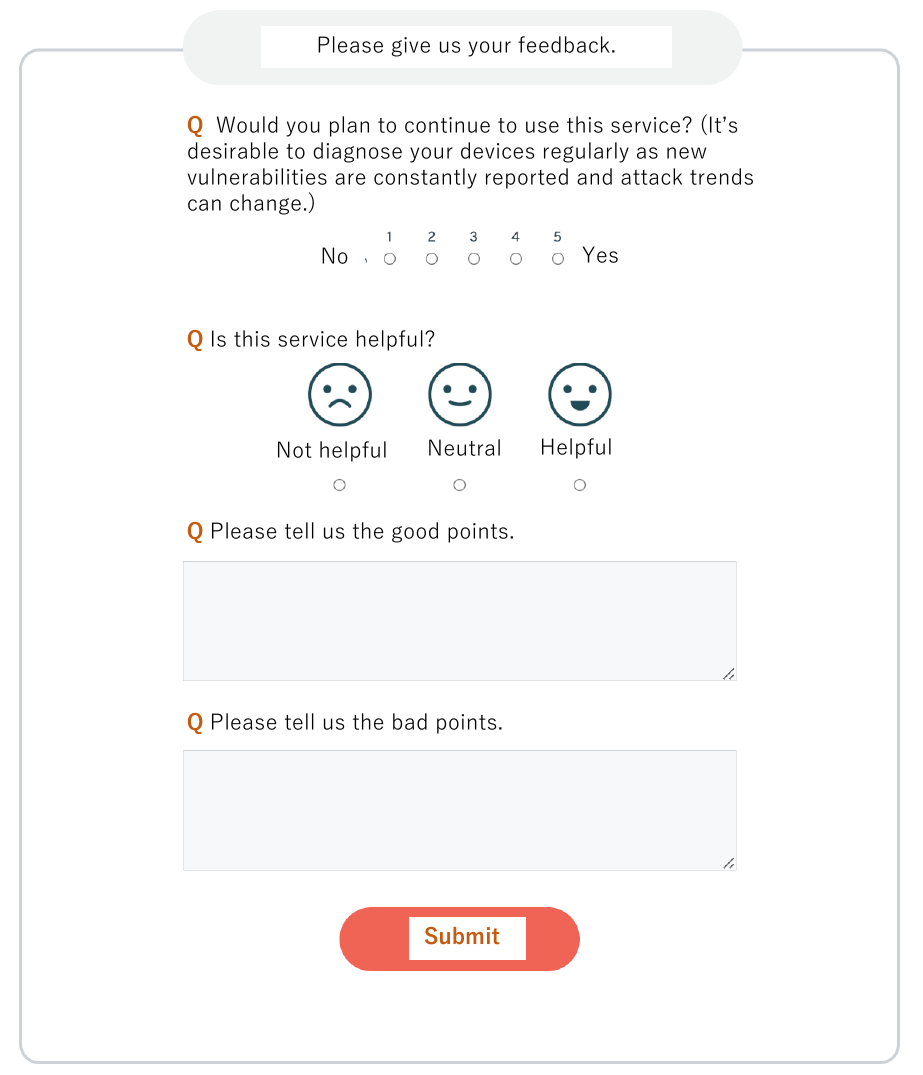}
    \caption{Questionnaire about engagement (Translated from Japanese)}
    \label{fig:questionnaire-about-engagement}
\end{figure}

\update{Figure~\ref{fig:questionnaire-about-engagement} shows questionnaire about engagement. We asked whether the users would like to continue using the service, whether the service was helpful, and good/bad points.}

\color{black}
\section{Details of notification results}
\label{appendix:details-of-notification-results}

Table~\ref{tab:num_of_detected_users_per_vulnerability} shows the number of users of each risk. 
Since a user may have multiple risks, the total number of risks exceeds the number of users.
Regarding known default IDs and credentials, we did not check whether the ID and password were actually set on the device due to ethical reasons. Instead, we checked whether the device model had known default IDs and credentials.

\update{Next, we analyze the results in terms of the device model~(Table~\ref{tab:idenfied-devices}).}
In the \updatenumY{274}{283} users' cases 
\update{where device models of risky devices were identified,}
four types of devices were identified: routers~(\updatenumY{36}{38} models from 5 manufacturers), NAS~(\updatenumY{51}{52} models from 3 manufacturers), web cameras~(\updatenumY{14}{15} models from 3 manufacturers), and a fir wall~(1 model from 1 manufacturer). \updatenumY{40}{42} models of IoT devices have been detected as ``End of support'', and the range of the sale's year is 2007--2016. In detail, \updatenumY{22}{23} models of QNAP's NAS have been detected and the range is 2009--2015, \updatenumY{13}{14} models of NEC's router have been detected and the range is 2010--2016, and 5 models of Panasonic's web camera have got detected and the range is 2007--2012.

Table~\ref{tab:remediation-rate} shows the remediation rates of each risk.
There was a difference in remediation rates depending on the type of risk. While the remediation rate for ``End of support'' was high, the rate for ``Risky protocol~(Telnet)'' was low.

\begin{table}[h]
    \centering
    \caption{Identified risks\updatefigC}
    \footnotesize
    \label{tab:num_of_detected_users_per_vulnerability}
    \scalebox{1.0}{
    \begin{tabular}{ll@{\hskip5pt}l@{\hskip5pt}l@{\hskip5pt}l@{\hskip5pt}l@{\hskip5pt}}
        \\\toprule
        Risks & Router & NAS & Web Cam. & Firewall & Total \\ 
        \midrule
        Malware infection$^*$ & - & - & - & - & 171\\ \midrule
        Known default ID & 98 & 48 & 6 & 0 & 154\\
        Old firmware & 47 & 24 & 1 & 41 & 113\\
        Risky proto.(Telnet)$^*$ & - & - & - & - & 121\\
        Known default cred. & 16 & 55 & 14 & 0 & 102\\
        End of support & 31 & 41 & 13 & 0 & 86\\
        Weak default Wi-Fi pass. & 37 & 0 & 0 & 0 & 37\\
        Known vulnerability & 26 & 0 & 0 & 3 & 29\\
        Admin password not set & 0 & 0 & 0 & 0 & 2\\
        No authentication & 0 & 0 & 0 & 0 & 1\\
        \bottomrule
    \end{tabular}
    }
    \scriptsize
    \flushleft{* For malware infection and Telnet detection, we did not identify device models.}
\end{table}

\color{black}

\label{sec:identified-devices}

\begin{table}[t]
    \centering
    \caption{Identified devices with risks\updatefigC 
   }
    \label{tab:idenfied-devices}
    \scriptsize
       \begin{tabular}{@{\hskip5pt}l@{\hskip5pt}l@{\hskip5pt}l@{\hskip5pt}l@{\hskip5pt}l@{\hskip5pt}l@{\hskip5pt}l@{\hskip5pt}}
        \\\toprule
        Dev. type & Manufacturer & Model~(Top 5) & Release year & \#Devices  \\ 
        \midrule
        Router & NEC & WR9500N & 2011 & 7   \\
               &     & WG1800HP2 & 2014 & 6 \\
               &     & WR8370N & 2010 & 4 \\
               &     & WG1800HP & 2013 & 4 \\
               &     & WR8600N & 2011 & 3 \\
               & BUFFALO & BHR-4GRV2 & 2014 & \updatenumC{36}{38}  \\
               &         & WSR-3200AX4S & 2020 & 5 \\
               &         & WXR-5700AX7S & 2020 & 4 \\
               &         & WXR-1900DHP3 & 2017 & 4 \\
               &         & WZR-300HP & 2012 & 2 \\
               & NETGEAR & RAX80 & 2019 & 3 \\
               &         & R8000P & 2017 & 1 \\
               & ELECOM & WRC-1167GST2 & 2018 & 1 \\
               & TP-Link & Archer C55 & 2016 & 2 \\
               &         & Archer C3150 & 2016 & 1 \\
               &         & Archer A10 & 2018 & 1 \\
        NAS & QNAP & TS-231P & 2016 & 11 \\
            &      & TS-220 & 2013 & 9\\
            &      & TS-230 & 2020 & 7 \\
            &      & TS-453Be & 2018 & 5 \\
            &      & TS-231+ & 2015 & 4 \\
            & NETGEAR & ReadyNAS & 2009--2017 & 1 \\
            & R.O.D & VS-2208Pro+ & 2017 & 3 \\
            &       & VS-2204Pro+ & 2017 & 1 \\
        Web Cam. & Panasonic & BB-SW175 & 2012 & 7 \\
                   &         & WV-SW458 & 2014 & 6 \\
                   &         & DG-SC385 & 2010 & 2 \\
                   &         & \update{BB-SW172} & 2012 & 2 \\
                   &         & \update{BB-ST165} & 2012 & 2 \\
                   & AXIS & M3044-V & 2016 & 4 \\
                   &      & \update{M2025-LE} & 2016 & 1 \\
                   & Canon & VB-C60 & 2008 & 1\\
        Firewall & Fortinet & Fortigate & 2002--2022 & 41 \\
        \bottomrule

    \end{tabular}
\end{table}

\begin{table}[t]
    \centering
    \caption{Remediation rate\updatefigC}
    \footnotesize
    \label{tab:remediation-rate}
    \#Confirmed remediations/\#Re-diagnosis
    \scalebox{0.85}{
    \begin{tabular}{@{\hskip2pt}l@{\hskip2pt}l@{\hskip2pt}l@{\hskip2pt}l@{\hskip2pt}l@{\hskip2pt}l@{\hskip2pt}}
        \\\toprule
        Risks & Router & NAS & Web Cam. & Firewall & Total \\ 
        \midrule
        Malware infection* & - & - & - & - & 59/67(88\%)\\ \midrule
        Known default ID & 18/38(47\%) & 7/14(50\%) & 1/2(50\%) & - & 27/55(49\%)\\
        Old firmware & 7/17(41\%) & 8/11(73\%) & 0/0 & 8/12(67\%) & 23/40(58\%)\\
        Risky proto.(Telnet) & - & - & - & - & 15/51(29\%)\\
        Known default cred. & 5/6(83\%) & 16/22(73\%) & 2/3(67\%) & - & 25/36(69\%)\\
        End of support & 6/10(60\%) & 15/17(88\%) & 3/3(100\%) & - & 25/31(81\%)\\
        Weak default Wi-Fi pass. & 8/13(62\%) & - & - & - & 8/13(62\%)\\
        Known vulnerability & 4/10(40\%) & - & - & 2/2(100\%) & 6/12(50\%)\\
        Admin password not set & - & - & - & - & 1/1(100\%)\\
        No authentication & - & - & - & - & 0/0\\
        \bottomrule
    \end{tabular}
    }
    \scriptsize
    \flushleft{* There would be false negatives caused by our detection method.}
\end{table}

\section{Cases of individual support}
\label{sec:individual-support}
We displayed our contact e-mail address on the top page of \amii~for support and got the following e-mails.

\begin{itemize}
    \item Inquiries about the usage of the service: \updatenumC{31}{34} cases. For example, the e-mails did not reach users due to spam filters, and users could not access the result page due to a disabled Cookie function of Web browsers. In these cases, our support team manually sent the users the diagnosis results by e-mail.
    
    \item Inquiries about the result of diagnosis or diagnosis methods: \updatenumC{20}{23} cases.
    Questions about the IP address shown on our service, and the IP address shown on the user's router are different. We noticed several reasons: there are multiple routers between the end users' environment and operators' network; and users checked IP addresses of their router assigned in a local area network (e.g. Wi-Fi at home) while the diagnosis is done on their global address. In these cases, we explained the diagnosis mechanism of our service.
    
   \item About the authenticity of our emails and website: 10 cases. We explained that our email and website were not spam or phishing. 

    \item Errors of the service: 9 cases.
    Users told us about high-load server errors caused by flash crowds due to the introduction by a TV program. We asked users to access the site again after a certain period of time.

    \item Questions about the security of IoT or general computer: 7 cases.
     
    \item Inquiries about the deletion of the collected information: \updatenumC{6}{9} cases. 
    We deleted the user information.

    \item Questions about the security of our service: 5 cases.
    
    \item \update{Support requests}~(Questions about how to take measures): \updatenumC{3}{4} cases.
    We supported the user by e-mail. 

    \item Others 3 cases. A report about a false positive, and two inquiries about changing the email address of a user.
\end{itemize}

We also got \updatenumC{two}{four} inquiries via the button to ask for support on the result page.
\begin{itemize}
\item A user whose device had been detected as malware infected inquired how to identify the detected device and take measures against it. We suggested following the router's manual to reboot it and update the firmware, with concrete instructions for finding the manual. There was no further interaction, and the user did not show any remediation.
\item A user whose device had been detected as malware infected inquired how to identify the detected device. We suggested almost the same advice as above and the user showed remediation later.
\item \update{A user whose device had a known default password. We suggested updating the password.}
\item \update{One user with a malware-infected device asked us whether the user needed to take action on the user's PC as well. Although it is unlikely that the malware targeting router could also infect the computer, we suggested a malware scan for the computer just in case.}

\end{itemize}

\section{Email template of diagnosis results (Translated from Japanese)}

\fbox{\begin{minipage}[h]{0.90\linewidth}
\small
Thank you for using \amii.\\
We completed the diagnosis of your router.\\

\lbrack Date of diagnosis \rbrack \\
2022/XX/XX XX:XX \\

\lbrack Page for diagnosis results and re-diagnosis \rbrack \\
Please access the following link to view the details of the inspection results.\\
If you wish to re-diagnose after taking measures, please use this link.\\

(A link to a URL of a diagnosis result)\\

If you do not recognize this e-mail, please discard this e-mail.
\\ \\
(Our contact information)

\end{minipage}}

\section{Answer length of questionnaire about engagement }
\begin{figure}[h]
    \centering
        \includegraphics[width=0.85\linewidth]{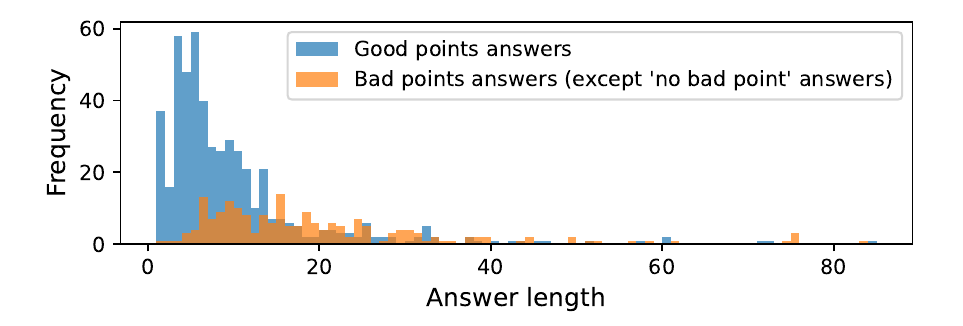}
    \caption{Histogram of answer length (\#English words)}
    \label{fig:hist-answer-length-english}
\end{figure}

\update{
Figure~\ref{fig:hist-answer-length-english} shows the word counts for the 500 responses translated into English. Regardless of the response length, codes were assigned based on response content.
}

\onecolumn

\section{Questionnaire items about barriers to take measures (Translated from Japanese)}
\label{sec:questionnaire-items}

\begin{figure}[h]
    \centering
    \includegraphics[width=0.69\linewidth]{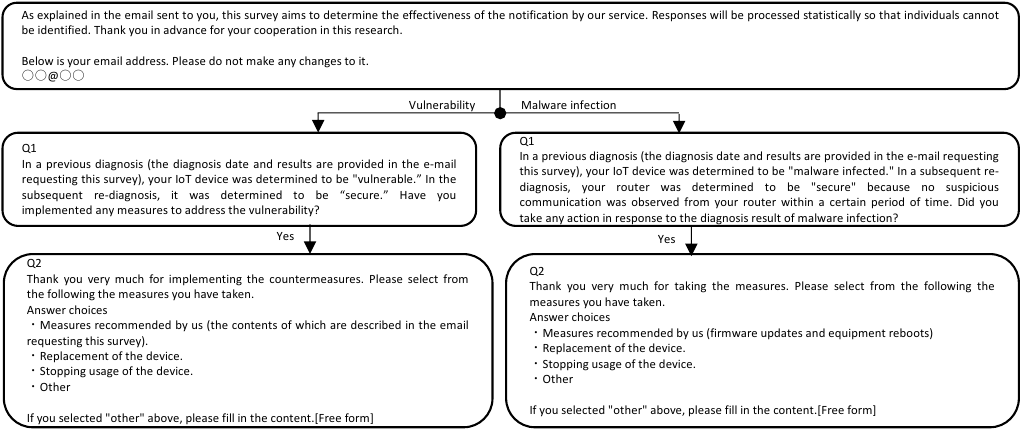}
    \caption{Questionnaire items for users with remediation confirmed}
    \label{fig:questionnaire_items_for_A}
\end{figure}

\begin{figure}[h]
    \centering
    \includegraphics[width=0.69\linewidth]{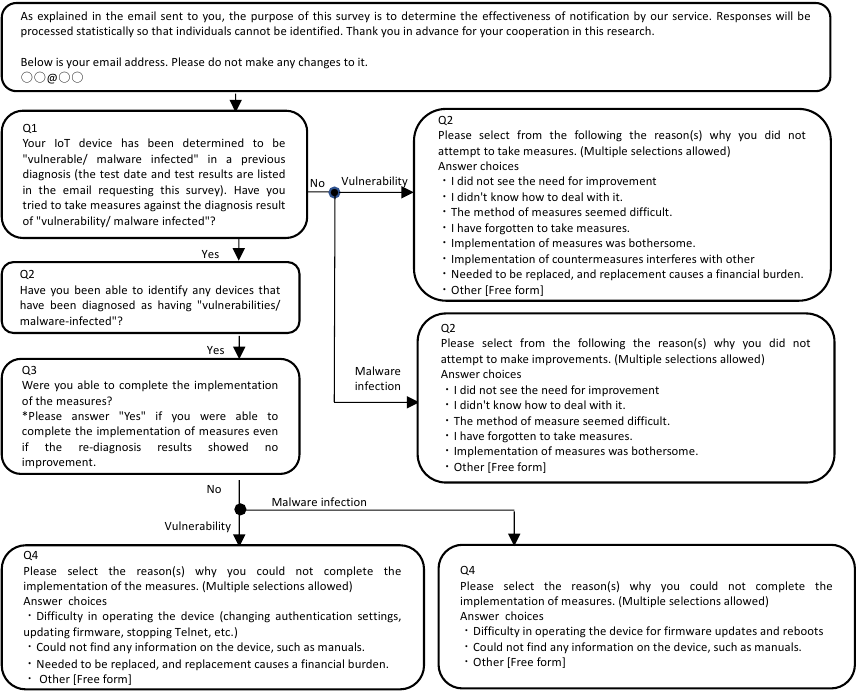}
    \caption{Questionnaire items for users with remediation not confirmed}
    \label{fig:questionnaire_items_for_B}
\end{figure}

\end{document}